# Future change in the solar wind and Central England temperature: implications for climate change attribution.

Ian R. Edmonds.

(Physics Department, Queensland University of Technology, Brisbane, Australia (Retired))

iredmonds@aapt.net.au

**Abstract** The recent increase in global temperature is attributed by the IPPC to anthropogenic global warming, (AGW), with a minor role for natural trends in temperature due to solar activity and volcanism. The IPPC estimates natural temperature (NAT) from climate models and attributes the difference from recent recorded temperature to AGW. This paper uses the temperature record to assess if trends in temperature are due to NAT or AGW effects. The method requires long records like the 362-year Central England temperature (CET) record. The CET was divided into a 262 year-long early part when only NAT was significant, and a 100 year-long later part. The early part was decomposed into eight components in the spectral range 15 to 257 years and the components were forward projected to the next 100 years. The projected NAT replicated the recorded cooling from 1950 to 1980 and the rapid increase from 1980 to 2010, indicating that the recent strong 50-year trend in CET was primarily NAT. Based on the small difference between the projected NAT and the recorded CET a minor role was attributed to AGW and a climate sensitivity to $CO_2$ doubling, $\Delta T_{2CO2}$ = 0.7 +/- 0.2 C, was estimated. Components, at 514 and 1028 years, were derived from the CET record providing a means for validation of long projections against proxy records of past temperature. Future projection of combined NAT and AGW indicated a cooling of CET by 0.5 C from now to year 2060 before AGW becomes dominant. The possible cause of an imminent decrease in CET was explored by applying the same method of component estimation to temperature data from Melbourne, Australia (MET) and the geomagnetic aa index, a proxy for the solar wind. Comparing cyclic variations of the aa index and the CET and MET data indicated a complex relationship with the strong recent increase in CET and MET lagging the increase in the solar wind by ~15 years.

**Highlights**

Novel method used to decompose the CET and aa index into cycles

Forward projection of past CET onto recent CET demonstrates recent CET is primarily natural

Observational determination of CET sensitivity to $CO_2$ doubling = 0.7 +/- 0.2 K

Exploration of the complex lagged relationship between the solar wind and temperature

## 1. Introduction.

The prospect of catastrophic global warming, as a result of increasing anthropogenic emissions, has resulted in the prediction of future climate change becoming a critically important area of science, IPCC (2013b). The variation in long temperature records such as the Central England temperature, (CET), is due to natural variations and, in more recent times, according to the IPCC, the effects of anthropogenic emissions, principally $CO_2$ and $SO_2$, IPPC (2013a). The radiative forcing due to $CO_2$ is given by $\Delta F = 5.35\ln(C/C_0)$ Wm$^{-2}$, Myhre et al (1998). The change in temperature due to $CO_2$ forcing is given by

$$\Delta T = \lambda \Delta T = \lambda 5.35\ln(C/C_0) \tag{1}$$

where C is the current $CO_2$ concentration, $C_0$ the pre industrial $CO_2$ concentration, and $\lambda$ is a sensitivity factor or attribution factor, the value of which depends on how much of recent temperature change is attributed to the forcing due to $CO_2$ emissions, (IPPC 2013a, Hansen et al 2011). Natural temperature (NAT) variation is due to solar irradiance variation, volcanism, oceanic oscillations such as the Pacific Decadal Oscillation, and other factors such as cloud changes due to changes in cosmic ray flux, Dorman (2021). If $\lambda$ is known equation 1 can be used to predict future AGW for various $CO_2$ emission scenarios, (Stott et al 2006a, IPCC 2013b). However, due to uncertain feedback effects in the climate system $\lambda$ is difficult to calculate from first principles. To estimate $\lambda$ computer models of NAT are made and the difference between recorded temperature and the modelled NAT is attributed to different effects including AGW, (Stott and Kettleborough 2002, Stott et al 2006a, Knutson et al 2006, Stott et al 2006b). The temperature change attributed to $CO_2$, $\Delta T$, and the current $CO_2$ concentration, C, are used with equation 1 to determine a value for the sensitivity, $\lambda$. If all the temperature increase since 1850, 1.2 C, is attributed to the increase of $CO_2$ from 300 to 400 ppm, the sensitivity factor $\lambda$ = 0.8 K/Wm$^{-2}$. This is a very simplified description of attribution based on climate models, currently a complicated, vast, and expensive scientific endeavour. Attribution is the critical process and relies on accurate modelling of NAT.

The CET, extending over 362 years from 1659 to 2022, Figure 1, is the longest instrumental record of temperature, (Manley 1974, Parker et al 1992). The linear trend in CET is 0.28 °C/century and the variance of CET is high; the standard deviation of the detrended CET is 0.6 °C. The high variance in CET presents challenges in attributing the contribution of AGW to CET. One approach to attribution that can be used with long temperature records is to compare historic trends in NAT with recent trends in NAT. If a recent trend is uniquely high, the trend could be regarded as unlikely to be due to NAT and, on that basis, attributed to AGW. Karoly and Stott (2006) applied this approach to the CET and attributed the recent 50-year trend in CET to AGW. The approach is valid only if the entire temperature record is considered. Karoly and Stott (2006) omitted the first part of the CET record that, according to recent work, shows a 50-year trend, 1690 to 1740, of 0.35 °C/decade, significantly stronger than the recent 50-year trend, 0.27 °C/decade, (Tung and Zhou 2013, Gonzalez-Hildalgo 2020). If all the recent increase in CET is attributed to AGW, $\lambda$ would be 0.8 K/Wm$^{-2}$ and the projected increase in CET would follow the dotted line in Figure 1, a projection predicting an increase of CET of about 5 °C from present levels by 2100 and a climate sensitivity due to doubling of $CO_2$, $\Delta T_{2CO2}$ = 3.3 °C.

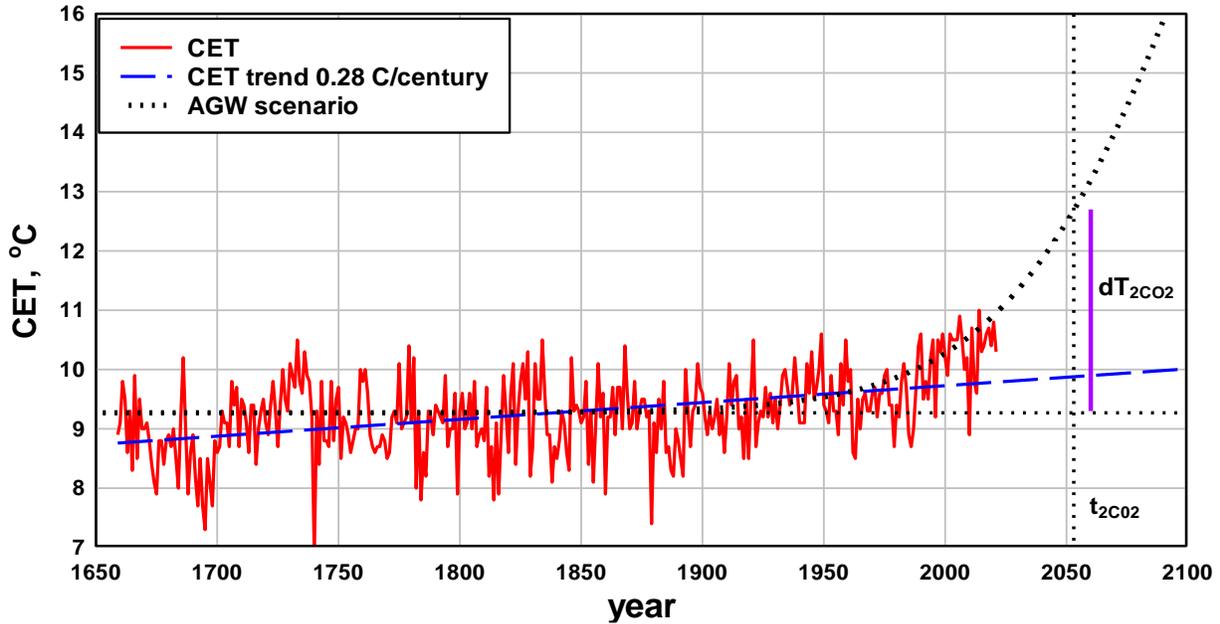

Figure 1. CET 1659 to 2021 and the linear trend in CET of 0.028 °C/century. If the recent positive trend in CET between 1950 and 2000 is attributed entirely to AGW an AGW scenario can be fitted to this trend and used to make a forward projection of CET, the dotted line in the graph. The fit is obtained by assuming an exponential increase in $CO_2$ concentration, C, between $C_0$ = 300 ppm in 1850 and C = 400 ppm in 2010 and a logarithmic dependence of temperature on $CO_2$ concentration, $\Delta T = \lambda \Delta F = \lambda 5.35 \ln(C/C_0)$. The temperature increase obtained is constrained to fit the recent trend in CET by adjusting the temperature to $CO_2$ sensitivity factor, $\lambda$, to the value 0.80 K/Wm$^{-2}$, corresponding to a $\Delta T_{2CO2}$ of 3.3 °C on $CO_2$ concentration doubling from an 1850 level of 300 ppm to 600 ppm by the time of $CO_2$ doubling, $t_{2CO2}$, marked by the vertical reference line.

The CET experienced a positive 50-year trend in the past, evident in Figure 1, between 1690 to 1740, significantly larger than the recent trend, (Tung and Zhou 2013, Gonzalez-Hildalgo et al 2020), so the question of attribution arises: Is the recent high trend in CET primarily attributable to AGW or is it primarily attributable to the same natural effect that resulted in the earlier and larger trend? If the recent 50-year trend in CET was primarily natural the fitting of an AGW scenario to the recent trend as in Figure 1 would overestimate future temperature change: a viewpoint supported by several recent analyses of various temperature records, (Wu et al 2011, Loehle and Scafetta 2011, Humlum et al 2011, Tung and Zhou 2013, Abbot and Marohasy 2017).

Clearly, the strong variations in CET before 1900, (Tung and Zhou 2013, Gonzalez-Hidalgo 2020), were due to natural effects. The earlier variation could be due to natural variation such as volcanism and/or due to some complex superposition of natural cycles originating from solar irradiance and cosmic ray variation, de Jager (2005). There is considerable evidence of cycles in the CET, (Plaut et al 1995, Baliunas et al 1997, Tung and Zhou 2013), as well as evidence of long- term persistence in the CET, Gonzalez-Hidalgo (2020). Spectral analysis of other long temperature records provides evidence of cyclic behaviour, (Humlum et al 2011, Humlum et al 2012, Scafetta 2010, 2021). This paper builds on this evidence by decomposing the CET into several components; each component derived from one of a series of spectral bands covering the period range 15 to 1000 years. Section 3 outlines how the extraction of components with period longer than the CET record length was achieved with a new method of spectral decomposition. Cycles were fitted to the components and the cycles were projected and superposed to obtain back and forward projections of the CET. In section 4 the validity of this approach was assessed by back projecting the CET and comparing with proxy records of past temperature. In Section 5 the same process of component and cycle identification was applied to the CET data in the time range 1659 to 1921 and a forward projection to the 100 years from 1921 to 2021 was made and compared with the actual CET data for that interval, in particular the time interval 1950 to 2010 that exhibits the recent strong 50-year trend in CET, Figure 1. This forward projection was used to attribute the recent temperature variation between natural and secular effects and assess climate sensitivity. In the last part of Section 5 the change in temperature was compared with the change in aa index, a proxy for the solar wind. Section 6 discusses the validity of IPCC projections of catastrophic temperature increase in the light of the projection of this paper of imminent temperature decrease. Section 7 is a conclusion.

## 2. Data sources

Annual mean Central England temperature, (Manley 1974, Parker et al 1992) is available at https://www.metoffice.gov.uk/hadobs/hadcet/data/meantemp_monthly_totals.txt. Graph of the CET anomaly is available at https://www.metoffice.gov.uk/hadobs/hadcet/ . The Melbourne daily maximum temperature was obtained from http://www.bom.gov.au/climate/data/ for the Melbourne Regional Office data 1855 to 2013 and from the Olympic Park data 2013 to 2023. The geomagnetic aa index, 1868 – 2021, can be downloaded from https://geomagnetism.ga.gov.au/geomagnetic-indices/aa-index

## 3. Method

**3.1 The spectral content of CET.** The spectral content of the detrended CET, Figure 2, was assessed by standard Fourier analysis. There are narrow peaks in the short period range, 15 to 57 years, but, due to limited resolution of the Fourier analysis, in the longer period range peaks are replaced by bands; one broad band between 60 and 100 years and the other between 200 and 400 years. It was noticed that the periods of some of the narrow peaks corresponded closely to harmonics of the Uranus-Neptune conjunction period, $T_{UN}$ = 171.4 years. For example, 57 = $T_{UN}/3$, 34 = $T_{UN}/5$, 24 = $T_{UN}/7$, and 15 = $T_{UN}/11$. In the absence of other criteria, the spectral range was divided into bands with centre periods, T, based on harmonic and simple factors of $T_{UN}$, i.e., selecting bands with centre periods, T, by using the relation $nT = mT_{UN}$ where n and m are small integers. The basis for selecting periods based on $T_{UN}$ is strong evidence, as outlined in detail in Appendix A3, that both solar activity and temperature records are dominated by cycles with periods close to harmonics and sub harmonics of $T_{UN}$. Further, back projection of the CET, as developed later in this paper, reproduces both the coarse and fine detail of proxy temperature records over the last two millennia when the back projection is based on decomposing CET into cycles based on harmonics and sub harmonics of $T_{UN}$, c.f. Section 5.4.

The centre periods used correspond to the labelled reference lines in Figure 2 with two exceptions. Spectral power at the period $T_{UN}$ is absent from the CET record for reasons outside the scope of the present article to discuss; but see McCracken et al (2014). The labelled period at 114 years, T = $2T_{UN}/3$, was omitted for reasons outlined in Appendix A1.

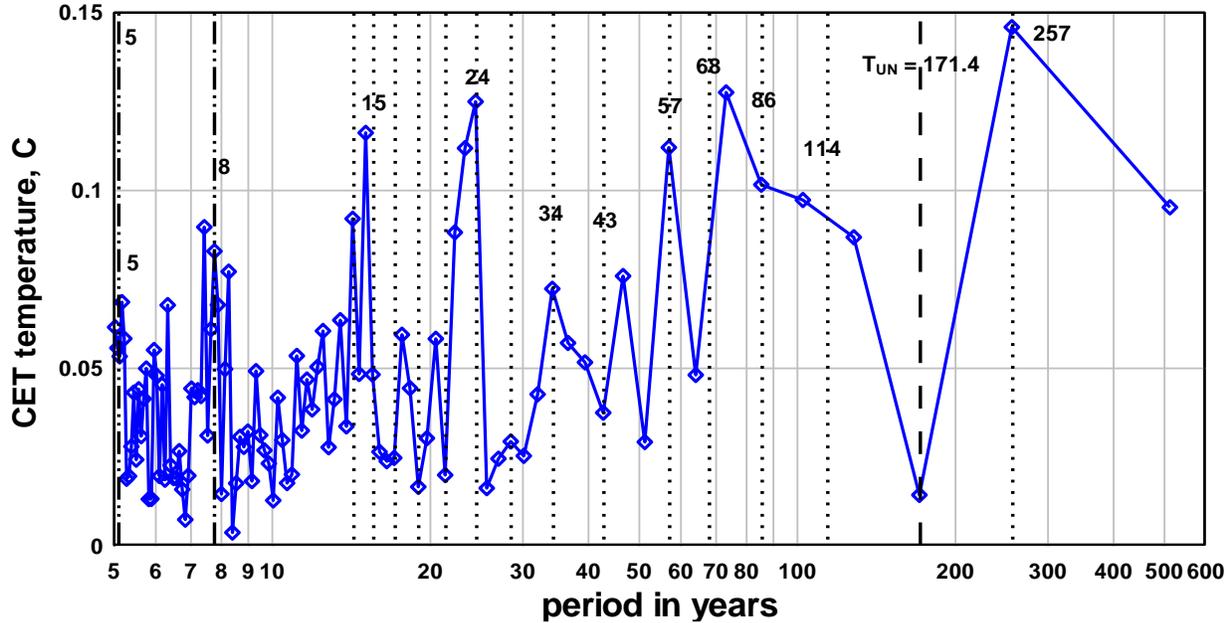

**Figure 2.** Periodogram of detrended CET obtained by padded Fast Fourier Transform. The dotted reference lines are at periods corresponding to harmonics or simple factors of $T_{UN}$ = 171.4 years. e.g., 15 = $T_{UN}/11$, 24= $T_{UN}/7$, 34 = $T_{UN}/5$, 43 = $T_{UN}/4$, 57 = $T_{UN}/3$, 68 = $2T_{UN}/5$, 86 = $T_{UN}/2$, and 257 = $3T_{UN}/2$ years.

**3.2 Method of decomposition of the detrended CET into cycles.**

The detrended CET was decomposed into components based on eight frequency bands of centre frequency $k/T_{UN}$ where k = 2/3, 2, 5/2, 3, 4, 5, 7, and 11. For example, for k = 3, the band centre frequency is $3/T_{UN}$ = 0.0175 $yr^{-1}$, corresponding to the period 57.13 years. The method of decomposition is illustrated in Figure 3 where the 24.5-year component, k = 7, is obtained. A Press band reject filter, Press et al (2007), as implemented in the DPlot application, is applied to the detrended CET. To reject the 24.5-year component the centre frequency of the Press filter was set to 1/24.5 = 0.0408 $yr^{-1}$ and the filter bandwidth set to 10% of the centre frequency, 0.004 $yr^{-1}$. The filtered CET is then subtracted from the unfiltered CET yielding the 24.5-year component, indicated by the bold blue line in Figure 3. The time variation of this component is consistent with the time variation of the 24-year component in CET obtained by the more conventional method of continuous wavelet spectrum analysis, Tung and Zhou (2013). The eight components obtained are shown in Figure 4. As each component is decomposed from a frequency band the components, as expected, show significant variation in amplitude and phase. However, due to the narrow filter bandwidths the largest phase shift of any component over the 362-year record was a 1/2 cycle shift for the k = 11, 15-year component. The average phase shift of the eight components is 1/6 of a cycle indicating that fitting a cycle to each component, apart from the k = 11, 15-year component, is a very good approximation. A comparison of the sum of the eight components with a seven-year running average version of detrended CET, Figure 5, indicates that the sum of the eight components closely reconstitutes the detrended CET. The correlation coefficient, r = 0.92. In the following CET will refer to the seven-year running average of the annual mean CET.

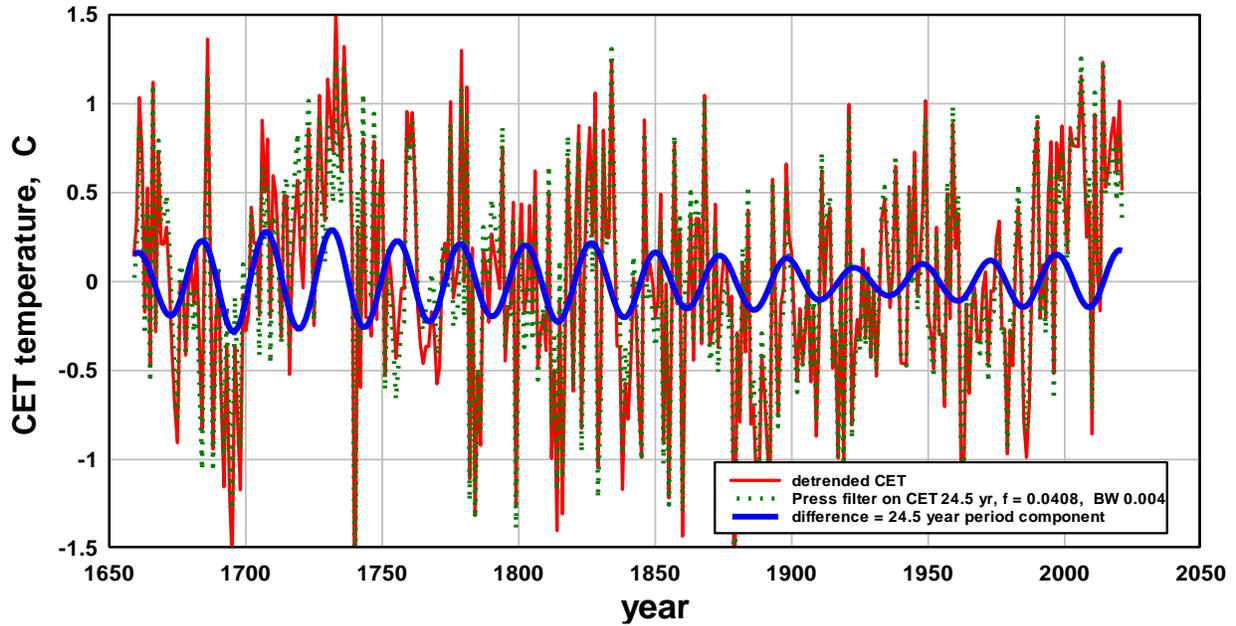

**Figure 3.** Illustrates the novel decomposition method. To obtain the component of CET in the band with centre period $T_{UN}/7 = 24.5$ years the detrended CET, red line, is filtered using a Press band reject filter of centre frequency 0.0408 yr$^{-1}$ with 10% bandwidth, 0.004 yr$^{-1}$, green dots. The difference between the original CET and the filtered CET yields the 24.5-year period component of CET, blue bold line.

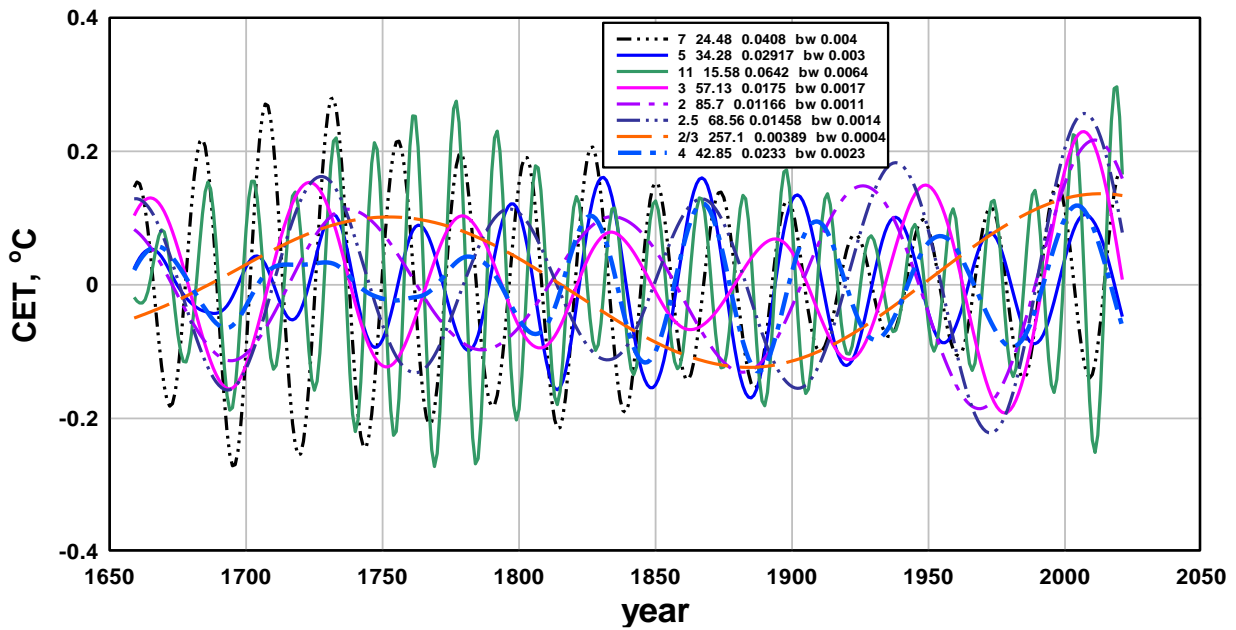

**Figure 4.** The eight components decomposed from the detrended CET using the method of Press band reject filtration outlined in the text. The centre frequency and 10% bandwidth of each frequency band are indicated. The average amplitude of the components is ~0.15 °C.

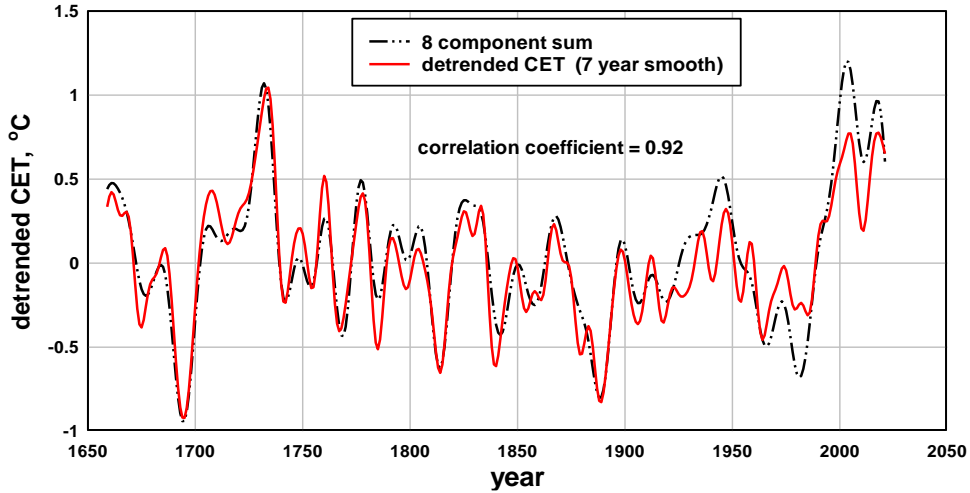

Figure 5. Adding the eight components in Figure 4 closely reconstitutes the seven-year running average of the CET. Of particular interest are the high 50-year trend 1690 to 1740, the increase from 1900 to 1950, the decrease from 1950 to 1980, the rapid increase 1980 to 2010 and the hiatus 2000 to 2021.

To project CET backward or forward it is necessary to approximate components with sinusoids. The approximation was obtained by fitting sinusoids of the form $0.15\cos(2\pi k_i(t - t_i)/T_{UN})$ to each of the components in Figure 4. Here $k_i/T_{UN}$ defines the frequency of the ith sinusoid. For example, when $k_i = 3$, $k_i/T_{UN} = 0.0175$ yr$^{-1}$ corresponding to period 57.1 years. The year $t_i$ defines the phase of the sinusoid. The fit is obtained by adjusting $t_i$ to maximise the correlation coefficient between the component and the fitted cycle. The resulting eight cycles, the sum of the cycles, and the detrended CET are shown in Figure 6. The approximation of the detrended CET with the eight constant amplitude cycles, Figure 6, reduces the correlation coefficient to $r = 0.68$, Figure 6. However, the major features, the high warming trend 1690 to 1740, the cooling 1950 to 1980, and high warming trend 1980 to 2010 are retained.

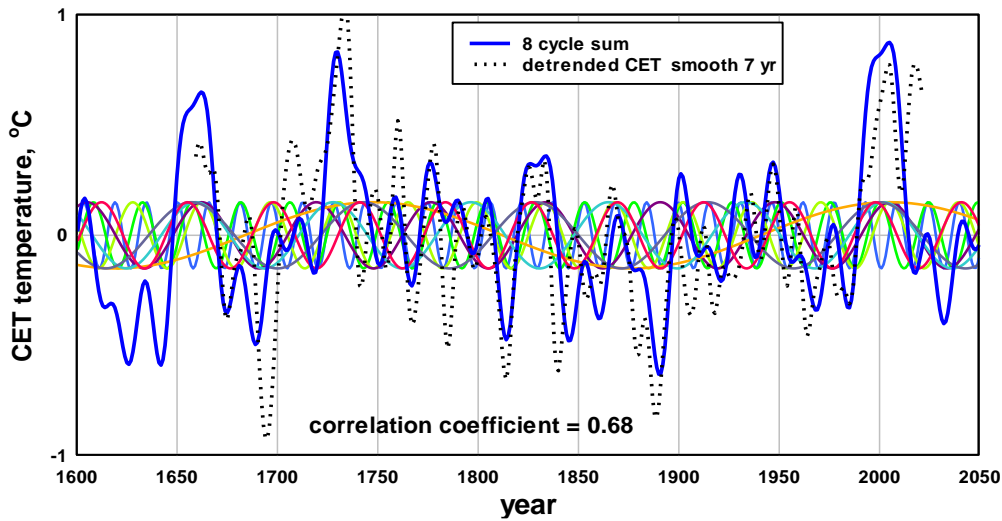

Figure 6. Shows the constant amplitude sinusoidal cycles fitted to the components of Figure 4. The sum of the eight cycles, blue bold line, and the detrended CET, dots, are also shown. The approximation of the components by cycles results in a moderately good approximation to the detrended CET, correlation coefficient $r = 0.68$.

**3.3 The variation in CET due to long period cycles.** It is evident that the linear trend of CET, 0.28 °C per century, Figure 1, cannot extend indefinitely into the future or extend indefinitely into the past and must be due to long term cyclic variations in temperature imposed on the CET. Fourier analysis of the 362-year CET record, Figure 2, has insufficient resolution to uncover very long cycles. However, proxy records of solar activity and related quantities are available over much longer intervals (Castagnoli et al 1992, Rigozo et al 2010, Abreu et al 2012, Scafetta 2012, McCracken et al 2013, McCracken et al 2014). Fourier analysis of the records revealed the presence of long period cycles in solar activity several of which occur at periods close to harmonic and sub harmonic periods of $T_{UN}$. The ~1000-year Eddy cycle, (Eddy 1976, Ma 2007, Zhao et al 2020) is close in period to $6T_{UN}$ = 1028 years. Castagnoli et al (1992) identified six major components at 1100, 690, 500, 340, 250, and 90 years, closely corresponding to periods at factors of $T_{UN}$, respectively, 6, 4, 3, 2, 3/2, 1/2. One of the major periodicities obtained from proxy cosmic ray data by McCracken et al (2013) occurred at 510 years, close to $3T_{UN}$, and two of the periodicities, 976 years and 1126 years, are both close to $6T_{UN}$. Scafetta (2012) found strong components in proxy total solar irradiance data at 499 years, ~$3T_{UN}$, and at 978 years, ~$6T_{UN}$. Ludecke and Weiss (2017) used worldwide temperature proxies to construct a mean global temperature over the last 2000 years. The harmonic analysis of the mean global temperature showed strongest components at periods at 1000 years, ~$6T_{UN}$, and at 460 years, ~ $3T_{UN}$. Humlum et al (2011) used components at 1190 years, ~$6T_{UN}$, and at 560 years, ~$3T_{UN}$, to reconstitute the Central Greenland temperature as obtained from the GSIP2 ice core. Abbot (2021) applied spectral analysis to eight proxy temperature records for the northern hemisphere and found the dominant periodicities in the millennial and centennial time range were ~1000 years and ~500 years. Based on this observational evidence the same method of band reject filtering as outlined above was applied to the CET record to find the components of CET at $3T_{UN}$ (514 years) and at $6T_{UN}$ (1028 years). Cycles of the same amplitude as previously, 0.15 °C, were then fitted to the components by the same correlation method as outlined previously. The sum of the two long period cycles obtained is shown in Figure 7. Also shown in Figure 7 is the sum of the two components, the CET anomaly, and the linear trend of the CET anomaly. It is apparent that the linear trend of 0.28 °C/century in CET obtained in Figure 1 is partly due to the long period cycles and partly due to the short-term temperature depression around 1700 and the short-term temperature enhancement around 2000. The two long-term cycles were added to the eight-cycle approximation of CET, Figure 6, and the sum of the ten cycles was obtained using

$$CET = 0.15 \sum_i \cos\left(\frac{2\pi k_i(t-t_i)}{T_{UN}}\right) \ + \ 9.27 \quad °C \tag{2}$$

Here i = 1 …10. The ith value of $k_i$ is, in sequence, 11, 7, 5, 4, 3, 5/2, 2, 2/3, 1/3, 1/6, and the ith value of $t_i$ is, in sequence, 1993, 2000, 2005, 1998, 2005, 2002, 2000, 2005, 2000, and 2130 years. The periods, in sequence, are 15.6, 24.5, 34.3, 42.8, 57.1, 68.6, 85.7, 257, 514, and 1028 years. The constant temperature, 9.27 °C, is the average value of CET 1659 to 2021. With the two long period cycles included in equation 2 the correlation coefficient between the ten-cycle approximation and the CET record is increased to r = 0.78, Figure 8.

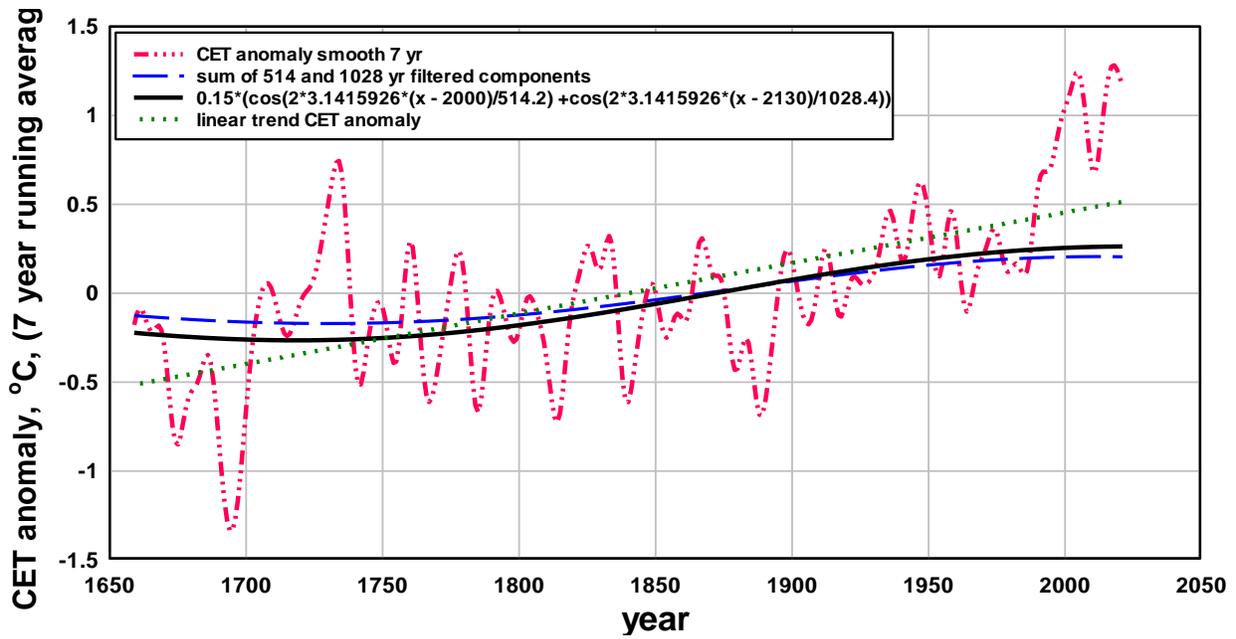

Figure 7. Shows the seven-year running average of CET anomaly, i.e., CET minus the 9.273 °C average. Also shown the sum of the 514-year and 1028-year components, the sum of the 514 and the 1028-year period cycles approximating the components, and the linear trend of the CET anomaly.

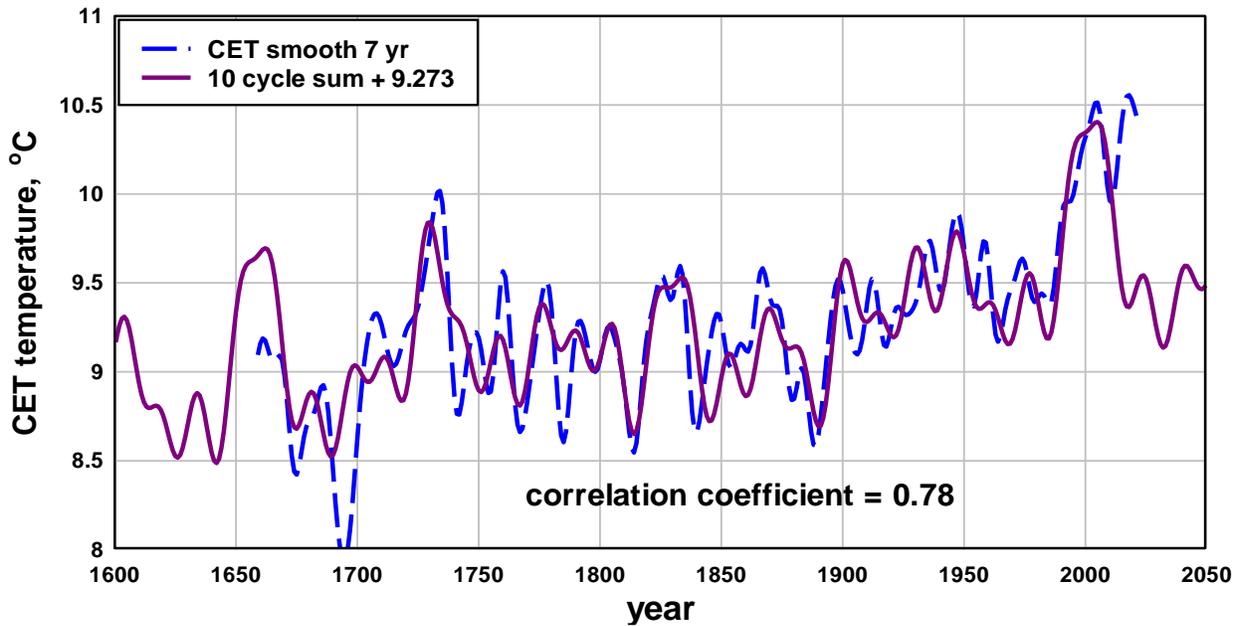

Figure 8. The ten-cycle approximation, equation 2, closely reconstitutes the CET. The correlation coefficient in the time range 1659 to 2021 is 0.78. The high 50-year trend, 1690 to 1740, the 0.5 °C rise, 1900 to 1950, the 0.5 °C decrease, 1950 to 1980, and the 1 °C increase from 1980 to 2010 are accurately reproduced by the ten-cycle approximation.

## 4. Backward and forward projection of the cyclic content of CET using equation 2.

The forward projection of the cyclic content of CET using equation 2 to the year 2500 is shown in Figure 9. There is short term variation in temperature due to the seven cycles in the period range 15 to 257 years superimposed on a longer-term variation due to the two cycles of period 514 and 1028 years. The shorter-term variation repeats at intervals of 343 years and gives rise to the rapid temperature changes evident at 1650, 2000 and 2350 years.

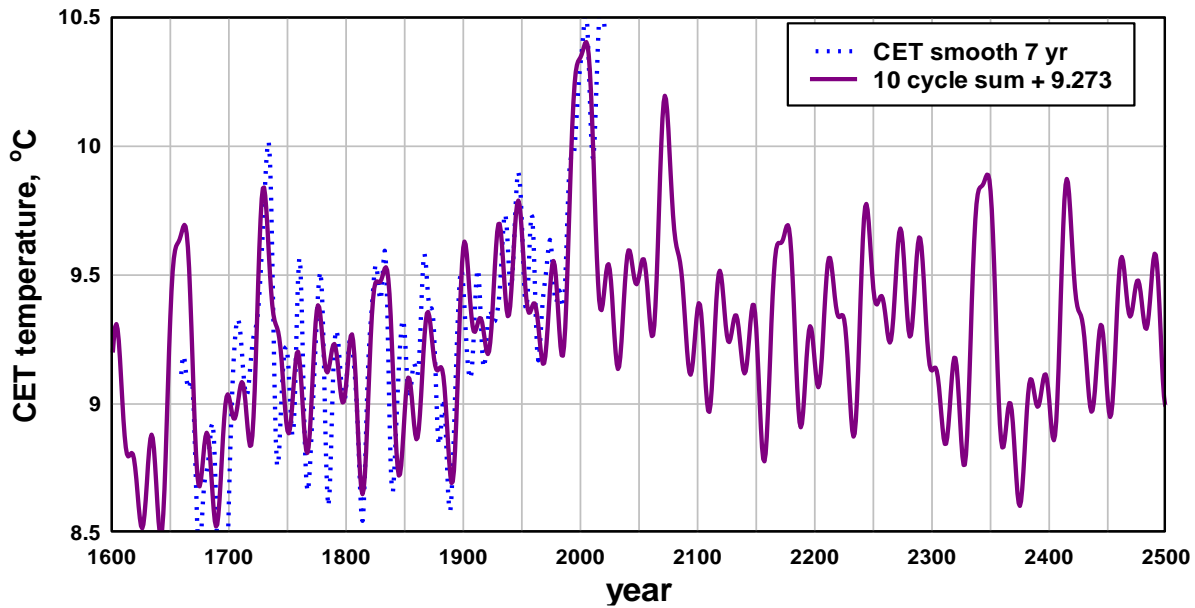

**Figure 9. The forward projection of the cyclic content of CET as given by equation 2. Note that the pattern of strong short-term variation, evident near years 1650, 2000 and 2350 repeats at intervals of 343 years The short-term variation is superposed on the longer-term cycles illustrated in Figure 10.**

The longer-term cyclic variation of CET can be best appreciated by calculating equation 2 with only the 514-year and 1028-year cycles, i = 9 and 10, included, as in Figure 10, to cover both back and forward projection. Figure 10 shows that the long-term cycles of period $3T_{UN}$ and $6T_{UN}$ years combine to produce the broad swings in CET between the Warm Periods and the Little Ice Ages that occurred during the past two millennia, as labelled in Figure 10 with the common terminology relating to the events. The multi-proxy reconstruction of temperature by Ljungqvist (2016) closely matches the major temperature excursions of this back projection providing strong confidence in the projection capability of equation 2.

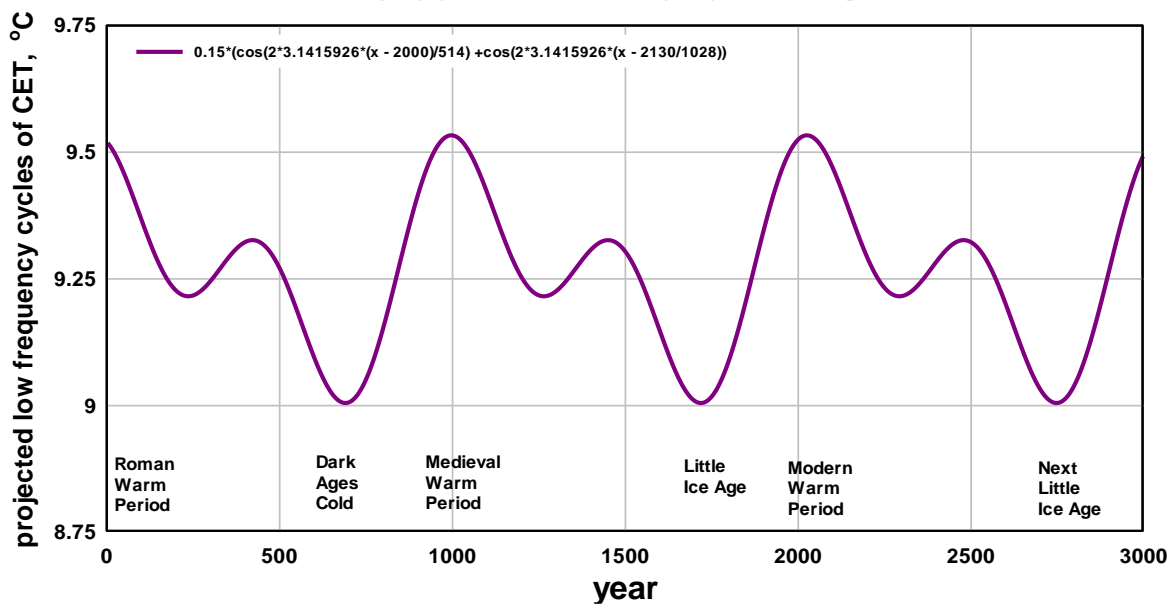

**Figure 10. The longer-term cyclic component of CET accurately reproduces the warm and cold periods identified in historical and proxy records of temperature.**

**5. Attribution of temperature change between secular and natural effects**

Attribution of recent trends in temperature to different forcings is the critical problem of climate science. The previous section has shown the detrended CET can be closely approximated by eight components and approximated moderately well, correlation coefficient r = 0.68, by eight cycles. Decomposing a record into components and cycles does provide a view as to how various cycles combine to produce extreme excursions and strong trends. For example, it is evident in Figures 4 and 6 that the strong trends in the early part and in the recent part of the CET record are due to the 57.1, 68.5 and 86.7-year components repeating an in-phase condition after an interval of about 340 years. However, the positive interference of cycles in different parts of the record is not in itself useful for attribution as the cycles may have resulted from different effects on temperature at different times in the record; for example, volcanism in the early part of the record and AGW in the recent part of the record. The only unambiguous conclusion is that the strong trends in the early part of the record were not caused by AGW.

Attribution based on climate models of NAT is problematic due to the large uncertainty in the NAT output of different models, e.g., (Karoly et al 2003, Karoly and Braganza 2005). Therefore, there is considerable interest in methods of attribution that avoid using climate models, Hegerl and Zwiers (2011). Wu et al (2011) used ensemble empirical mode decomposition to decompose the global mean surface temperature into a low frequency oscillation and a high frequency, approximately 65-year period, oscillation, and estimated about one third of late twentieth century warming was due to NAT. Another method is to decompose the temperature record into high and low frequency components using low pass filtering. The components of temperature are then compared with components of forcing decomposed from the same frequency range. For example, Tung and Zhou (2013) used low pass filtering to show the 50-to-90-year period signal in CET varied coherently with both the global mean temperature and with the Atlantic Multidecadal Oscillation (AMO) and attributed much of the recent CET variation to the AMO.

**5.1 Attribution by forward projection of early CET onto later CET.** As discussed in Section 4, approximating components with sinusoids provides for projection beyond the temperature record. Projection is especially useful to separate cyclic effects from secular or "one off" effects. It is generally accepted that AGW is the result of secular forcings that become strong after 1950 and that temperature change before 1950 is due primarily to natural effects, (Karoly et al 2003, IPCC 2013, Figure TS.19, IPCC Report 2021, Figure SPM.1, SPM.2b). It follows that, if NAT is cyclic, and the cyclic content can be accurately decomposed from temperature data before 1950, the cycles can be projected forward to assess the NAT after 1950. Loehle and Scafetta (2011) used this projection method, decomposing a 20-year cycle and a 60-year cycle from the global surface temperature record in the interval 1850 to 1950 and projecting the two cycles forward to 2010 to compare with the temperature record 1950 to 2010. From the comparison Loehle and Scafetta (2011) attributed 60% of the warming observed since 1970 to NAT. Humlum et al (2011) projected three cycles of period 71.7, 24.9, and 15.3 years, obtained from the Svalbard temperature record forward to 2035 and concluded that the late 20$^{th}$ century warming in Svalbard is not going to continue for the next 20 -25 years. Abbot and Marohasy (2017) fitted cycles to records of proxy temperature before 1830 and projected the cycles forward for comparison with the proxy records in the interval from 1880 to the present; finding that the increase in temperature over the last 100 years can be largely attributed to NAT. The results in section 3 of this paper indicate that the detrended CET can be accurately characterized by eight cycles covering the spectral range between 15 and 257 years, a spectral range relevant for the assessment of the 50-year trend in CET in the interval between 1950 and 2010.

The CET record was divided into a 262 year long earlier part, 1659 to 1920, and a 100 year long later part, 1921 to 2021. The method for deriving an eight-cycle simulation of the detrended CET as outlined above was applied to the 262 year long data to obtain components and the approximate cycles in the same eight spectral bands between 15 years and 257 years. The periods of the cycles and the constant amplitude, 0.15 $^{o}$C, of the cycles was retained, however, the phases, $t_i$, of the eight cycles obtained differ and were, in sequence i = 1 to 8, as follows: 1994, 1999, 2004, 1996, 2012, 1997, 1994, and 2016. The projected CET obtained with these values in equation 2 is shown as the red full line in Figure 11, where the projected CET is compared with (1), the actual CET, and (2), the projected CET based on the entire, 362 year long, data record. Clearly, the projection from the early part of the record replicates, reasonably accurately, the later, 1921 to 2021, part of the record including the warming between 1930 and 1950, the cooling between 1950 and 1980, and the strong warming from year 1980 to year 2010. It is noticeable in Figure 11 that the second very narrow warming peak in CET at 2018 was not reconstituted from the 1659 to 1921 data or from the 1659 to 2021 data. This is due to the 15-year period cycle approximation being half a period out of phase with the 15-year component towards the end of the record as discussed in Appendix A2. This phase shift between cycle and component is an inevitable consequence of approximating a narrow band component with a single period cycle. The two forward projections to 2200, the one based on the earlier part of the record and the one based on the entire record are, as evident in Figure 11, closely similar. The correlation coefficient between the projected CET and the actual CET in the time interval 1921

– 2021 is r = 0.55 indicating that the forward projection of the eight cycles reproduces CET in this interval moderately well. The correlation coefficient of the projection and CET in the 1921 to 2010 interval is r = 0.79. The result provides confidence in the forward projection capability of the method and confidence that the warming, 1920 to 1950, the cooling, 1950 to 1980, and the strong warming, 1980 to 2010, is primarily natural. The type of natural forcing that leads to the strong warming trend from 1950 to 2010 is not defined by this method. However, the possible candidates are solar activity and/or volcanism

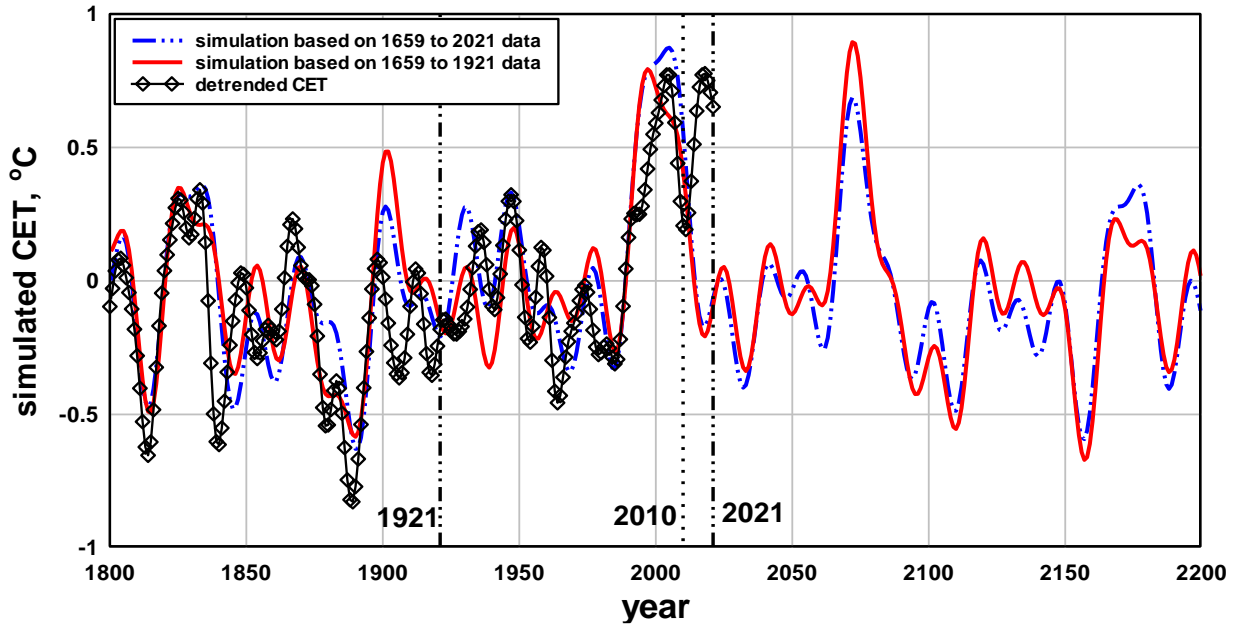

Figure 11. The cyclic variation of CET derived from the CET record 1659 to 2021 compared with the detrended CET, line with diamond symbols. The vertical reference lines indicate the time interval 1921 to 2021 and the time interval 1921 to 2010. In the 1921 to 2021 interval the correlation coefficient between the CET forward projection and the detrended CET is r = 0.55. In the interval 1921 to 2010 the correlation coefficient is r = 0.79. The difference in correlation is due to phase shift of the 15-year cycle relative to the 15-year component as discussed in Appendix A2.

**5.2 Climate sensitivity to AGW**. Previous work, for example Karoly and Stott (2006), attributed all the recent 50-year trend in CET to AGW with the implication that the sensitivity of temperature to $CO_2$ is high and the projected temperature increase due to increasing $CO_2$ is high. The forward projection of this high sensitivity scenario is illustrated in Figure 12

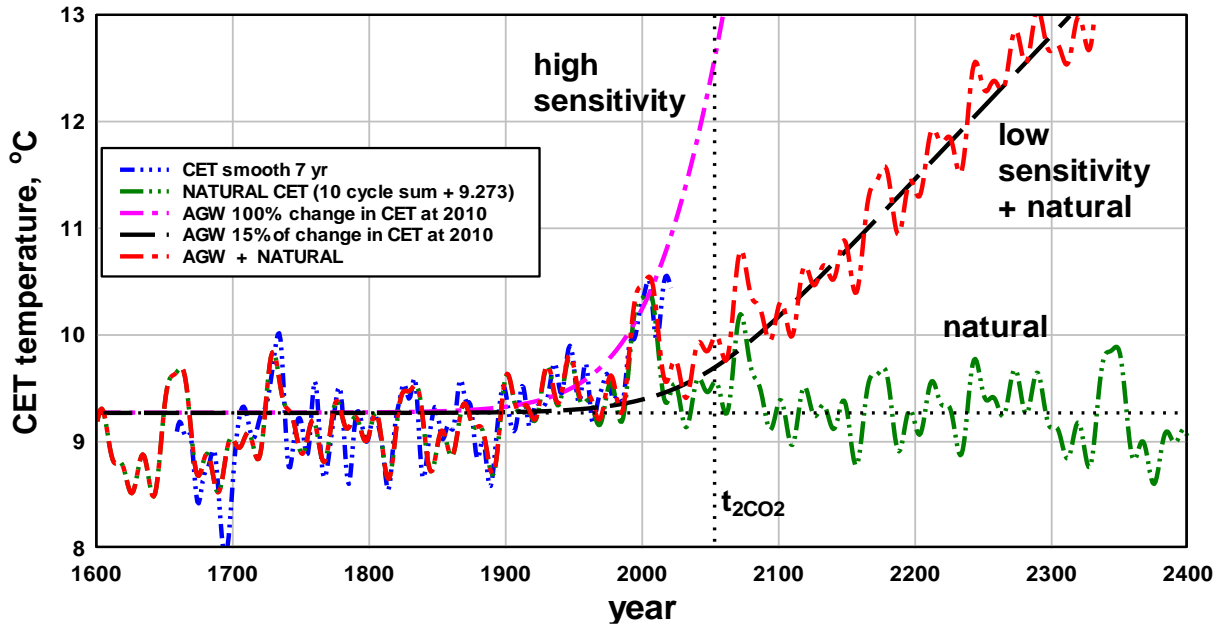

Figure 12. The seven-year running average of CET is shown by the blue broken curve. When the increase in CET from 1950 to 2000 is attributed entirely to AGW and the $CO_2$ concentration continues to increase at the present rate, a temperature increase of about five degrees at year 2100 can be projected. The projected natural CET, calculated by equation 2, is also shown, as the green broken line that closely overlaps the CET before 2021. With the 50-year trend from 1950 to 2000 attributed primarily to natural variation a much lower climate sensitivity, $\lambda$ = 0.10 K/Wm$^{-2}$, rather than $\lambda$ = 0.87 K/Wm$^{-2}$, is appropriate. With $\lambda$ = 0.10 K/Wm$^{-2}$ the $CO_2$ contribution to CET is shown by the black long-dash line. The forward projection of CET due to combined AGW forcing and natural forcing of CET is shown by the broken red line. Reading off the graph at $t_{2CO2}$, for the low sensitivity curve the $\Delta T_{2CO2}$ is now 0.7 K compared with $\Delta T_{2CO2}$ = 3.3 K for the high sensitivity curve, c.f. Figure 1.

The correspondence of the forward projection obtained from early CET data, before changes in $CO_2$ or $SO_2$ were significant, to the CET data 1921 to 2021 provides strong evidence that the recent 50-year trend in CET is primarily natural. However, estimating the small change in CET at year 2000 that could be attributable to AGW is subject to high uncertainty because the change in CET due to AGW is a small difference between two near equal values, the CET record itself and the projected natural contribution to CET, c.f. Figure 12. The low sensitivity projection in Figure 12 was based on estimating that, of the 1.2 °C increase in CET from 1850 to 2000, just 0.2 +/- 0.1 °C, or about 15% of the 1.2 °C increase at year 2000 is attributable to AGW. From the level of this projection at the time of $CO_2$ doubling, $t_{2CO2}$, the temperature change due to $CO_2$ doubling is $T_{2CO2}$ is 0.7 +/- 0.2 °C, c.f. Figure 12. Even with $CO_2$ concentration increasing at an exponential rate like the RCP8.5 scenario, the indication of the low sensitivity projection in Figure 12 is that, in the next few decades, CET will fall to a temperature about 0.7 °C below the current seven-year average CET of 10.5 C, and remain near that lower temperature, ~ 9.7 °C, until about 2070 when CET will briefly rise to slightly above the current temperature. The projection indicates that CET will not be consistently above present temperatures until 2150. The projected temperature change in this low sensitivity forward projection is lower than the RCP8.5 projections of climate models, (Stott and Kettleborough 2002, Stott et al 2006, IPCC 2013, Knutti et al 2017, Tollefson 2020) that resemble the high sensitivity projection in Figure 12. For example, the IPCC RCP8.5 temperature projection at year 2100 is 4 °C above present levels and at year 2300 is 8 °C above present levels, (IPCC 2013 Figure 12.5), whereas the low sensitivity temperature projection of Figure 12 at year 2100 is 0.5 °C below present levels and at year 2300 is 2 °C above present levels. While considerably different from IPCC projections the low sensitivity, $\Delta T_{2CO2}$ ~ 0.7 °C, projection should be viewed in the light of the recent hiatus in global temperature increase, Tung and Chen (2018), the decreasing estimates of climate sensitivity over time, Gervais (2016), the low climate sensitivity estimates based on other observational determinations, e.g., $T_{2CO2}$ ~ 0.7 °C by Lindzen and Choi (2011), $T_{2CO2}$ ~ 0.7 °C by Abbot and Marohasy (2017), the low sensitivity obtained by other methods of climate modelling, e.g., $T_{2CO2}$ = 0.6 °C by Harde (2014), $T_{2CO2}$ = 0.6 °C, Coe et al (2021), in the light of projections of imminent decrease in solar activity, (Clilverd et al 2006, Steinhilber and Beer 2013, Velasco Herrera et al 2015, Yndestad and Solheim 2016, Matthes et al 2017, Velasco Herrera et al 2021), the recent fall of the heliospheric magnetic field, c.f. Figure 14, and imminent reduction in global warming, (Loehle and Scafetta 2011, Omrani et al 2022)

It is noted that the projections in Figure 12 are based on an exponential increase in concentration of $CO_2$, C(t), given by the relation C(t) = $C_0$ + exp(0.028(t – $t_0$)). The increase approximates the Mauna Loa $CO_2$ data record and the RCP 8.5 scenario of $CO_2$ increase, Tollefson (2020). There are numerous scenarios for long-term $CO_2$ concentration increase, some more plausible than others, (Hansen et al 2013, Moss et al 2010, Tollefson 2020). However, on current trends, a $CO_2$ concentration of 500 ppm by 2050 appears probable. Based on this anticipated $CO_2$ level and the projected $SO_2$ concentration decrease, Bellouin et al (2011), and using the high climate sensitivity of $\lambda$ ~ 0.9 K/Wm$^{-2}$, as proposed by, for example, (IPCC 2013, Figure 12.40, IPCC 2018, Ma et al 2022), the CET would be 2 °C higher by 2050, as in Figure 12. Based on the low climate sensitivity estimate of this paper the CET would be 0.5 °C lower by 2050. Thus, the CET in the next few decades should provide a clear indication of the relative validity of deterministic projections and climate modelling projections of climate sensitivity and future climate change.

**5.3 Attribution of recent temperature extremes.** It is obvious, for example from Figure 8 and 9, that the CET has increased by about 2 °C during the 340 years between the Little Ice Age around 1680 and the year 2021; and by about 1 °C from year 1850 to the year 2021. It is also clear from this paper that the increase is primarily natural and is largely due to the two long-term cycles coming into positive interference at around year 2000, c.f. Figure 10. Currently, there is concern that AGW is having a catastrophic effect on climate; for example, the statement by the UN Secretary-General at the 2022 United Nations Climate Change Conference: "We are on a highway to climate hell with our foot on the accelerator", The Guardian (2022); or "Assessing "Dangerous Climate Change"", Hansen et al (2013). There is evidence that daily maximum temperatures and daily minimum temperatures have increased recently, (Karoly and Braganza 2005, Fischer et al 2021, IPCC 2021 Figure SPM.3b). An edited collection of reports of this type of evidence appears in an annual series, now eight years long, published by the American Meteorological Society from 2013 to 2020 titled "Explaining Extreme Events of 2XXX From a Climate Perspective", e.g., Herring et al (2014). A very brief summary of the findings therein is that recent record high temperatures can be attributed to AGW. Examples of research specific to England are papers attributing the record high CET of 2014 to AGW, King et al (2015), and the paper by Christidis et al (2020) attributing the record temperatures in the UK in 2019 to AGW. CET has trended upward from a minimum during the Little Ice

Age, Figure 1, so it is to be expected that in recent times CET will be exceeding temperature levels and records set at previous times, Rahmstorf and Coumou (2011). It is clear, from the attribution analysis in Section 5.1, that, if the same type of analysis had been made on CET data as it existed in 1920, a forward projection from 1920 would have anticipated a spate of temperature records in the 2000 to 2020 interval. If the increase in CET, 1950 to 2010, is attributed primarily to AGW the rate at which new CET records occur would increase with time in proportion with the increasing trend in CET indicated by the extreme scenario in Figure 12, Rahmstorf and Coumou (2011). Under the low sensitivity scenario of Figure 12 the indication is that the spate of new record high temperatures in the first two decades of this century is a temporary phenomenon and new high temperature records will not occur until 2070.

**5.4 Attribution of past temperature extremes.** Proxy estimates of Earth's surface temperature during the past two millennia are of considerable and controversial interest with respect to attribution science; see Smerdon and Pollack (2016) for a review. As well as forward projection, equation 2 can be back projected. Because equation 2 is based on cycles at harmonic periods and simple factor periods of $T_{UN}$ and as $T_{UN}$ is a constant of the solar system, forward and back projection by equation 2 does not accumulate error due to period selection error and projections can be made over long time intervals. A back projection of CET to the year -500 AD is shown in Figure 13.

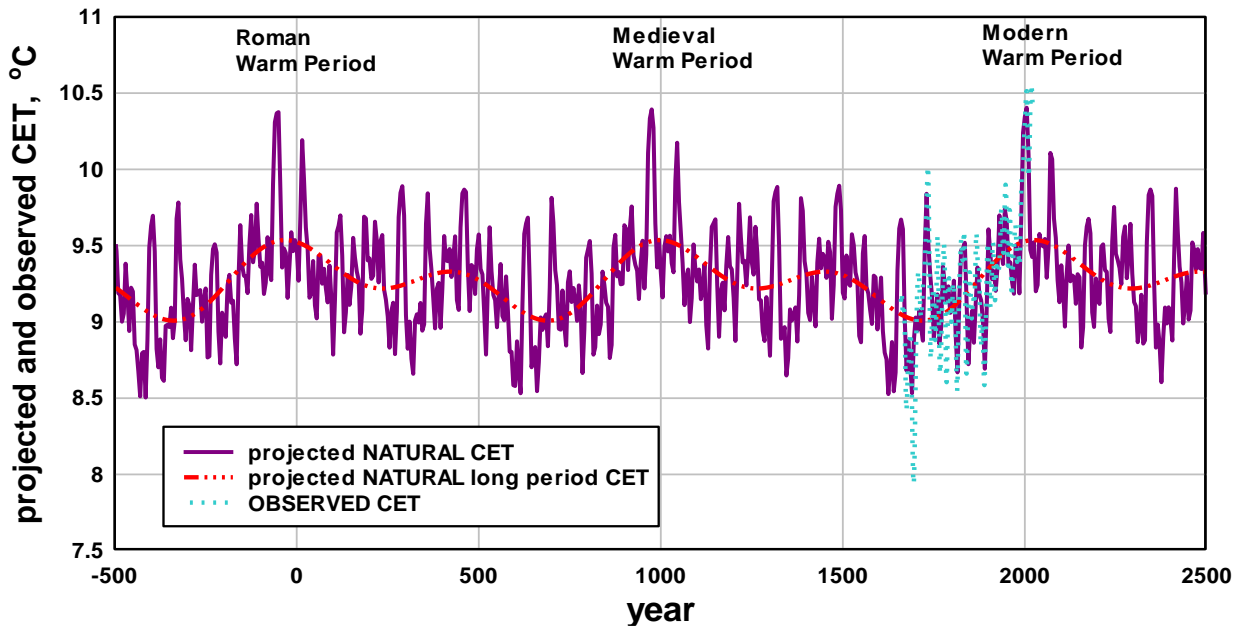

**Figure 13. The back projection of CET using equation 2. Also shown, broken red line, is the back projection with only the 514-year and 1028-year cycles in equation 2. It is evident that CET experiences extended Warm Periods at intervals of ~1028 years and extended cool periods, Little Ice Ages, also at intervals of 1028 years. Strong positive enhancements in CET occur at intervals of 343 years, e.g., at ~ year 0 and ~ year 343. Strong positive enhancements also occur within the extended cool periods, e.g., ~ year 680 and ~ year 1680, and within the extended warm periods, e.g., ~ year 0, ~ year 1000, and ~ year 2000.**

The extended warm periods around 0 AD, 1000 AD and 2000 AD in the projection coincide closely with the Roman, Medieval, and Modern Warm Periods identified from historical and proxy records of temperature, e.g. (Esper et al 2002, Mann et al 2008, Ljungqvist 2010, Buntgen et al 2011, IPPC 5$^{th}$ Assessment Report 2013, Figure 5.7, Smerdon and Pollack 2016, Ludecke and Weiss 2017), that indicate a slow 0.5 C swing between the Medieval Warm Period and the Little Ice Age with short term temperature variation increasing the swing to approximately 1 C. The back projection in Figure 13 shows two CET peaks, at year 975 and year 1045, that replicate the double temperature peak form of the Medieval Warm Period evident in new paleo-climate reconstructions, (Esper 2002, Moberg 2005, Mann et al 2008, IPPC 5$^{th}$ Assessment Report 2013, Figure 5.7). The back projection also shows two strong peaks at –55 and + 15 years that replicate the double temperature peak form of the Roman warm Period, Buntgen et al (2011). The coincidence provides further validation of the accuracy of equation 2 for long projections. Figure 13 shows that extended warm periods return at intervals close to $6T_{UN}$ = 1028 years and that strong positive and negative 50-year trends return at intervals close to $2T_{UN}$ = 343 years. For example, the strong 50-year trends in Figure 13 evident at around years 1000, 1350, 1690, and 2000 in Figure 13 are also evident in proxy temperature records; for example, at around year 1000 and 1350 in proxy records of Norther Hemisphere temperature, Mann et al (2008). It is interesting that the times of occurrence of strong 50-year trends match almost exactly the times of high spatial deviation between proxy temperatures from different regions of the Northern Hemisphere, Christiansen and Ljungqvist (2012). When the 343-year cycle of rapid CET variation overlaps with the extended warm periods in the projection the CET attains levels of 10.4 $^{o}$C, about one degree above the long-term average CET level and about two degrees above the minimum CET level. The indication from the back projection is that Central England is currently experiencing mean annual temperatures similar to the mean annual temperatures during the Medieval and the Roman Warm periods. Further, Figure 13 indicates that natural CET has increased by about 1.2 $^{o}$C during the last century and is now passing through the Modern Warm Period peak. It is well known that in a temperature record exhibiting a normally distributed temperature variation the fraction of temperatures that exceed some specified high temperature will increase exponentially as the mean temperature increases, (Hansen et al 2012, IPPC 5$^{th}$ Assessment Report 2013, Figure 1.8). The indication from Figure 13 is that the current spate of record temperatures in England, (King et al 2015, Christidis et al 2020), is due to a similar pattern of high temperatures and rapid temperature change imposed on long a term temperature cycle similar to the pattern that characterised temperature variation during the Roman and Medieval Warm Periods.

Temperature variation over the past two millennia as projected in Figure 13 and as observed in proxy temperature records, (Mann et al 2008, Ljungqvist 2010, Buntgen et al 2011, IPPC 5$^{th}$ Assessment Report 2013, Figure 5.7, Ludecke and Weiss 2017), presents a serious challenge to current attribution science. It is known from ice core records of trace gases, for example, Rubino et al (2019), that $CO_2$ concentration was essentially constant during the past two millennia and only began to increase from a level close to 280 ppm after 1800. Therefore, the temperature extremes of the Medieval Warm Period and the Little Ice Age along with the accompanying strong 50-year trends cannot be attributed to AGW and must be due to some other form of forcing, Shindell et al (2001). However, if the extremes and trends of the Medieval Warm Period and the Modern Warm Period are similar, as Figure 13 indicates, the attribution of the extremes and trends of the Modern Warm Period entirely to AGW is challenged. Mann et al (2009), suggest that "a better understanding of the influence of radiative forcing on large-scale climate dynamics should remain priorities as we work toward improving the regional credibility of climate model projections."

**5.5 Origin of recent natural variation in CET.** The IPCC attributes the recent rapid increase in temperature solely to AGW, (IPCC 2007, 2013, 2021). This paper attributes the recent increase in the CET primarily to natural effects. Attribution to natural effects was based on the forward projection of the cyclic content derived from the CET when only natural effects were significant. The projection to the recent 100-year interval coincided closely with the CET record in that interval indicating recent trends in CET were primarily due to natural effects. Attribution between natural warming and AGW by forward projection does not require specification of the natural effect, only that the cycles forward projected were obtained from a time in the record when AGW was insignificant. The two types of natural effect considered by the IPCC are volcanism and solar activity. As volcanism reduces temperature it is not relevant to the recent rapid increase in temperature. Increased solar activity would increase temperature. However, according to the IPCC, the effect of solar activity in recent times, is negligible.

The geomagnetic aa index, Figure 14, recorded since 1868, by magnetometers in England and Australia, is a proxy for the strength of the heliospheric magnetic field in the solar wind and is related to other forms of solar activity such as total solar irradiance, sunspot number and cosmic ray flux, Lockwood et al (1999)..

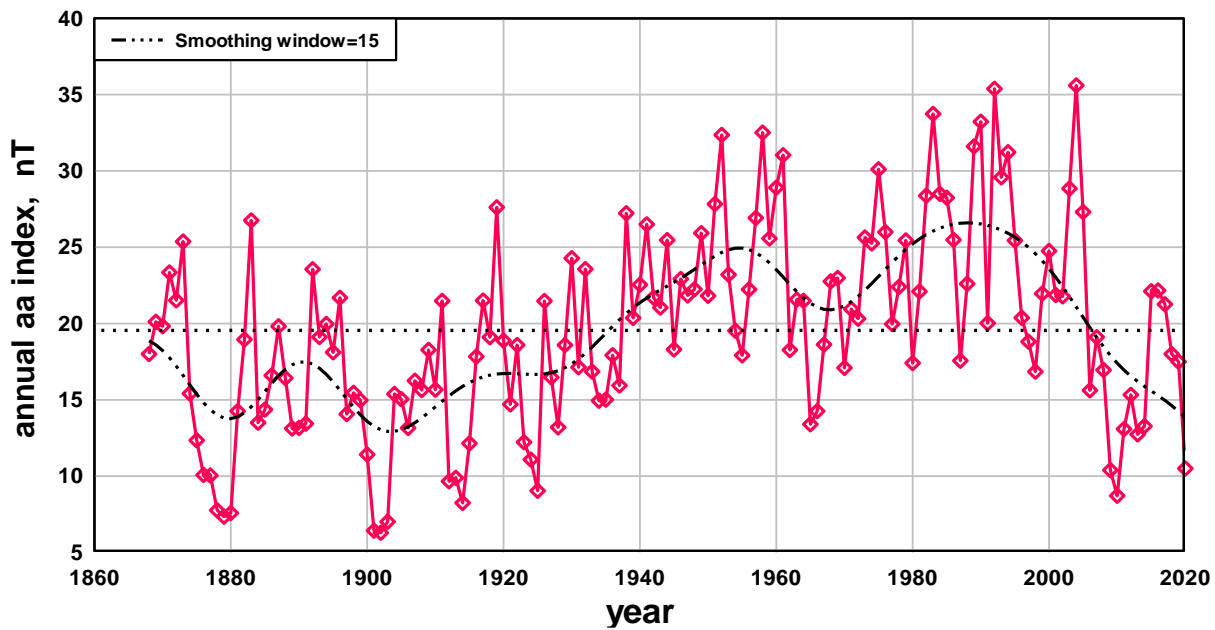

**Figure 14. Annual mean level of the geomagnetic aa index.**

The aa index roughly doubled between 1900 and 1990 due to the doubling of the solar magnetic flux emanating from the Sun, Lockwood et al (1999). This is the same interval during which the CET increased by about 0.6 C, c.f., Figure 1. The aa index fell sharply between 1990 and 2021 while the CET increased by another 0.6 C during the same interval, c.f. Figure 1. However, forward projection in this paper indicates an imminent fall in CET temperature suggesting that CET may follow a change in aa index after a significant lag. As the aa index record is long it is possible to apply the same decomposition into components method to the aa index as applied in section 4 to the CET record.

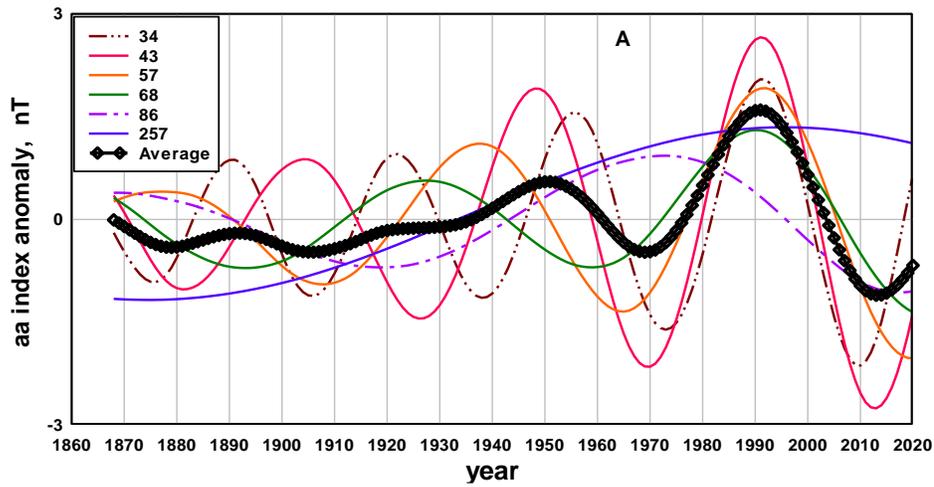

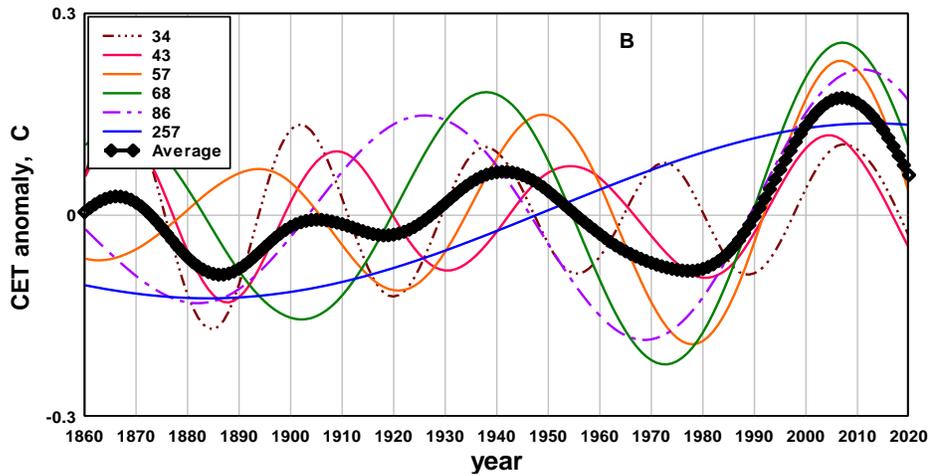

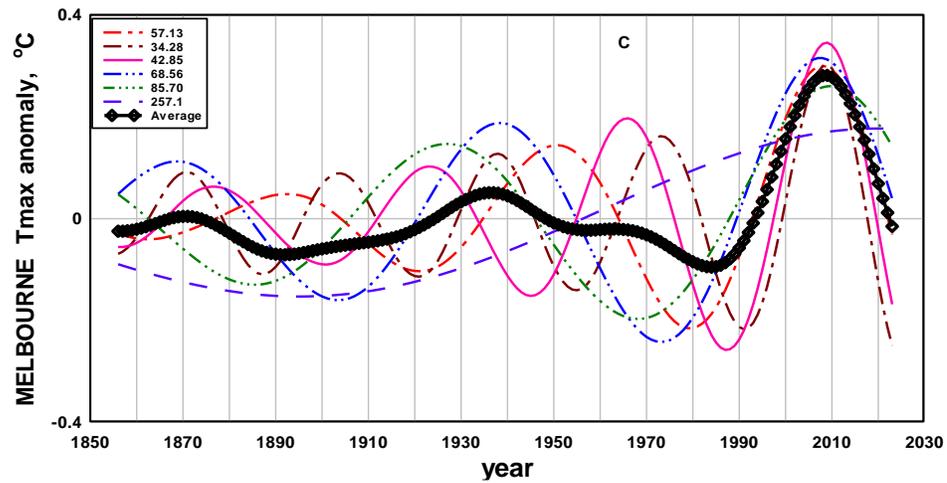

**Figure 15. The components of the aa index anomaly, (A), and the CET anomaly, (B), and the Melbourne $T_{MAX}$ anomaly, (C), derived from decomposition based on the periods 34, 43, 57, 68, 86 and 257 years. The average curves show that all components of the aa index positively interfere at year 1990 and all components of the CET and Melbourne $T_{MAX}$ positively interfere about 15 years later between year 2005 and 2010.**

As we are interested in trends over intervals of 50 years the aa index record was decomposed into components from frequency bands with centre periods at 34, 43, 57, 68, 86 and 257 years, Figure 15A. Also shown in Figure 15A is the average of the six aa index components. Figure 15B shows the six components derived from the CET anomaly for the same period bands, c.f., Figure 4. It is evident that a strong peak in the aa index, due to the positive interference of all six aa components, occurs in 1990, about 15 years before the strong peak in CET at year 2005 that is due to the positive interference of all six CET components. A similar result is obtained for the Melbourne $T_{MAX}$ anomaly (MET) with the six components coming into positive interference about 18 years after the peak in the aa index components. The average curves in Figure 15A, 15B and 15C show the same general trends of an increase from 1900 to 1940-1950, a decrease to year 1970-1980, an increase to peaks when the temperature peaks lag the aa index peak by about 15 years and then a sharp decrease from the peaks. The aa index additionally shows a minimum occurred at 2012. Provided the approximately 15-year lag of temperature relative to the solar wind is maintained in the next few decades, c.f., Figure 16, the implication is a minimum in CET and MET occurring at year 2027 - 2030. The correlation coefficient at zero lag between the two average curves in Figure 15A and 15B is r = - 0.06, suggesting an insignificant relationship. However, if the lag was considered the correlation would be high. Figures 15A, 15B and 15C indicate a strong but complex relationship between the cyclic variation of the aa index/solar wind and the cyclic variation of temperature. Close examination of Figures 15A, 15B and 15C shows that, aside from a lag that reaches a maximum lag of 15 years near year 2000, the 257, 68, 57 and 43-year components of aa index and temperature vary mainly in-phase. The 86 and 34-year components of aa index and temperature vary out-of-phase. This suggests that the response of temperature to the solar wind involves at least two internal oscillations, such as the Atlantic Multidecadal Oscillation, the North Atlantic Oscillation, the Pacific Decadal Oscillation, or the El Nino Southern Oscillation (Tung and Zhou 2013, Power et al 2017), that are forced by the solar wind and directly influence temperature. The results in Figure 15 indicate that the response of temperature to the solar wind is not amenable to simple regression analysis, as in de Jager et al (2010). For example, Figures 15A and 15B indicate that, while the aa index and CET anomalies, at zero lag, are negatively correlated in the 40-year interval between 1980 and 2020, the peak in CET at 2005 is likely a lagged response to the peak in solar wind/aa index at 1990. Lockwood and Frohlich (2007) used a method to analyse longer-term trends in solar activity and temperature between 1975 and 2007 and, noting the negative correlation, concluded that the rapid rise in temperature after 1985 could not be ascribed to solar variability. However, Lockwood and Frohlich (2007) did not consider the possibility that temperature change would significantly lag multidecadal solar forcing. It is relevant here that Li et al (2013) demonstrated that the North Atlantic Oscillation (NAO) is connected to the Northern hemisphere mean surface temperature over multidecadal time scales, of approximately 60-year period, with the NAO leading the surface temperature by 15-20 years and thus acting as a useful predictor of temperature change. Both the aa index/solar wind and the NAO peak at the same time, 1990, suggesting the two variables are related, c.f. (Gray et al 2013, Schaife et al 2013). Figures 15B and 15C indicate that the peak in regional temperature around 2010 is common at widely separated locations suggesting the influence of external forcing.

By fitting cycles to the aa index components in Figure 15A a six-cycle composite of the aa index can be obtained in an equation similar in form to equation 2:

$$aa = 1.5 \sum_i \cos\left(\frac{2\pi k_i (t - t_i)}{T_{UN}}\right) + 9.55 \quad nT \tag{3}$$

Here i = 1 …8. The ith value of $k_i$ is, in sequence, 5, 4, 3, 5/2, 2, 2/3, 1/3, 1/6, and the ith value of $t_i$ is, in sequence, 1991, 1991, 1992, 1992, 1967, 1999, 2000, and 2130 years. The periods, in sequence, are 34.3, 42.8, 57.1, 68.6, 85.7, 257, 514, and 1028 years. The constant, 9.55 nT, is the average value of the aa index 1868 to 2020. The $t_i$ values at i = 7, period 514 years, and at i = 8, period 1028 years, were not derived from the aa index record but are the same as the values obtained from the CET record, i.e., the phases of the 514 and 1028 period cycles were inferred to be the same as the phases of the corresponding cycles derived from the, much longer, CET record. Equation 2 and equation 3 can be used by interested readers to explore the time relationship between the aa index and CET. For example, the back and forward projection of the aa index and the CET, Figure 16, shows that the aa index mostly leads the CET but sometimes lags, e.g., 1900 to 1950, illustrating the complexity of the relation.

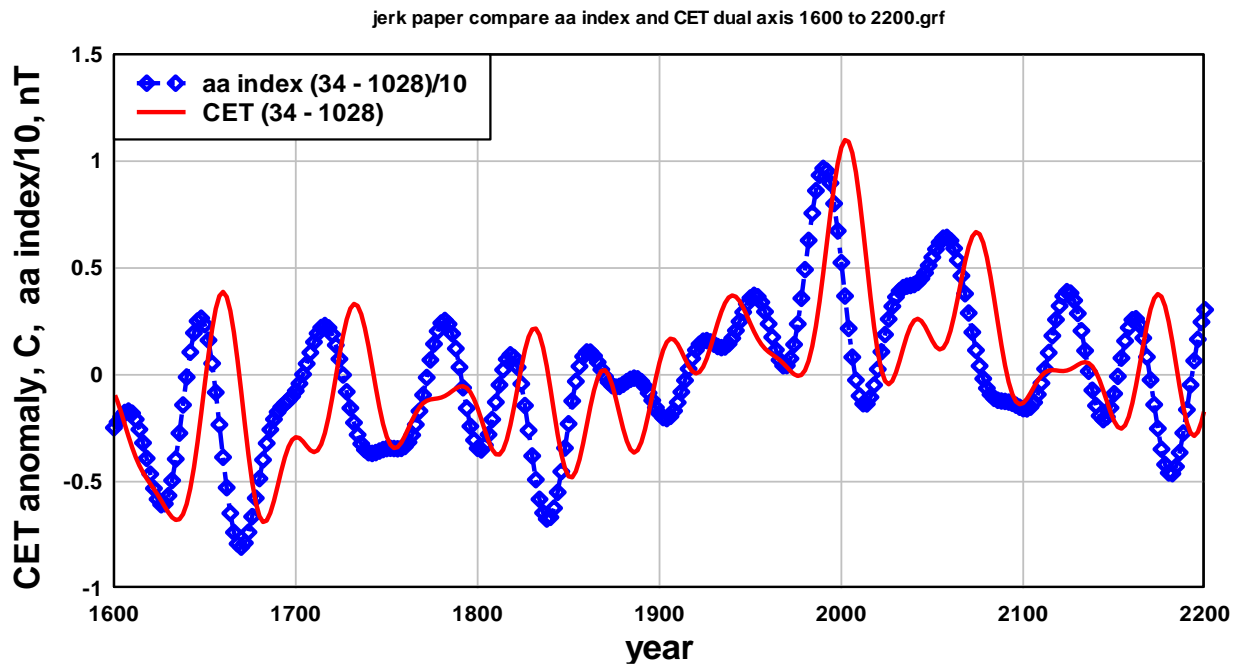

**Figure 16. Forward and back projection of the aa index anomaly (reduced by a factor of 10) and the CET anomaly using cycles with periods 34, 43, 57, 68, 86, 257, 514, and 1028 years in equations 2 and 3.**

The lag of the CET relative to the aa index/solar wind appears to be amplitude dependent, increasing to ~15 years when the aa index and the CET experience strong trends, e.g., near year 1650 and near year 2000. It seems likely that the delayed response of the CET to increase in the aa index is due to thermal inertia of the upper mixed layer of the ocean, Soldatenko (2022). Gray et al (2013) found that the sea surface temperature in the North Atlantic/European region lagged the 11-year solar cycle by 3 – 4 years or about 1/3 of a cycle and suggested the mechanism was solar forcing of the NAO, see Scaife et al (2013). See also Shindell et al (2001) who found "relatively small solar forcing may play a significant role in century-scale NH winter climate change".

## 6. Discussion

The changes in CET, 1659 to 2021, are quite complicated, Figure 1; however, the changes in temperature from 1900, when $CO_2$ emissions started to become significant is the current focus of climate science, (Hansen 2011, Hansen et al 2013, IPPC 2013, IPPC 2018). The changes in CET from 1900 are relatively simple; four trends in the temperature are evident, an increase of about 1 C from 1900 to 1950, a decrease of about 0.5 C from 1950 to 1980, an increase of about 1 C from 1980 to 2000, and a hiatus or a crest from 2000 to 2021. In climate modelling the increase 1900 to 1950 is attributed 80% to an increase in $CO_2$ and about 20% to an increase in solar irradiance; the decrease 1950 to 1980 is attributed to the combined effect of increases in $CO_2$, aerosols, and volcanism; the increase 1980 to 2000 is attributed primarily to the rapid increase in $CO_2$. In climate models the contribution from natural effects, specifically solar irradiance, to the increase between 1950 and 2000 is effectively zero, (Hansen et al 2011, IPCC 2021 Figure SPM.2b). The hiatus in CET, 2000 to 2021, presents a challenge for climate modelling and the hiatus generally is the subject of some controversy, e.g., Fyfe et al (2016). With the $CO_2$ concentration record continuing to increase exponentially at a rate equivalent to the RCP8.5 scenario and with the high climate sensitivity attributed to $CO_2$ for the 1950 to 2000 change in CET, an increase of about 0.3 C would be projected by climate models for the interval 2000 to 2021, IPCC (2013), and Figure 12. If the temperature hiatus continues, as projected for global temperature (Loehle and Scafetta 2011, Humlum et al 2011, Abbot and Marohasy 2017) or the temperature decreases in the next few decades as projected in this report and Li et al (2013) the challenge to the current climate modelling approach where the influence of external forcing is regarded as negligible would be severe.

This report indicates that the slow increase in CET, 1900 to 1950, was due to the positive phases of the 57, 68, and 86-year period components in CET beginning to overlap; the 1950 to 1980 decrease due to the negative phases of the same components nearly overlapping; and the sharp rise 1980 to 2000 due to the positive phases of the same components closely overlapping, Figure 4 and Figure 5. The hiatus, 2000 to 2021, would appear to be the transition interval between a lagged response to the increase in solar activity to a peak in 1990 and the lagged response of CET to a fall in solar activity from the peak in 1990. The projected decrease in CET between 2020 and 2050 indicated in Figure 12 and Figure 16, if it occurs, will be due to the negative phases of the 57, 68 and 86-year components in CET again closely overlapping before moving progressively out of phase. Thus, the observationally determined projection of CET indicates that the hiatus is likely a crest between the recent rise and the imminent decrease in CET. The result in Figure 15C for the Melbourne $T_{MAX}$ anomaly where the decrease in temperature from 2010 is evident provides support for the idea that the hiatus is more likely a crest.

By inference, any decreased climate sensitivity to $CO_2$ concentration requires an increased sensitivity to solar activity and the accompanying effects, such as variation in total solar irradiance, cosmic ray flux and cloud cover. The analysis in section 5.5 provides evidence that change in CET is a complex lagged response to variation in solar activity. Revised estimates of solar irradiance, Ergorova et al (2018), indicate the effect may be about six times higher than the estimates used by the IPCC. Forward projections of solar activity, as in Figure 16 and in other studies, (Rigozo et al 2010, Steinhilber and Beer 2013, Shepherd et al 2014), do predict a strong decrease in solar activity from the grand maximum around year 2000. It is interesting that decreasing solar activity from 2000 onwards has recently been the subject of global climate modelling; for example (Ineson et al 2015, Maycock et al 2015), and interesting that the Coupled Model Intercomparison Project (CMIP) has replaced the stationary-Sun scenario in CMIP5 with more realistic scenarios in CMIP6 that include solar activity falling to a grand solar minimum around 2100, Matthes et al (2017). A strong increase in future volcanic activity would also result in a decrease in temperature. There is evidence of an approximately a 50-70-year oscillation in volcanic forcing of temperature over the last two millennia e.g. (Mann et al 2021, Sun et al 2022). However, apparently, no projections of near-future volcanic activity have been made.

**7. Conclusion**

The analysis in this paper indicates that the CET is currently passing through a relatively short interval, about two decades long, of higher-than-average temperature that is primarily natural and is the result of several of the natural cycles that contribute to CET variation coming into constructive interference. This is the reason record high temperatures have recently been recorded in Central England. Comparison of the cyclic content of the CET with the cyclic content of the aa index indicates that forcing of the CET may be due to a complex interaction of solar activity, solar wind, cosmic ray flux and cloud cover with the global oscillatory systems such as the NAO that directly influence oceanic and atmospheric temperature. This report projects a decrease of about 0.5 °C in CET during the next few decades with CET temperature continuously exceeding present temperatures from 2100 onwards if $CO_2$ emissions continue to increase exponentially. The climate sensitivity, $\Delta T_{2CO2}$, is estimated to be 0.7 +/- 0.2 K.

**Appendix**

**A 1. Selection of component periods.** Basing the method of decomposing CET into cycles at periods based on simple factors of $T_{UN}$ is somewhat controversial. As pointed out earlier several of the shorter period peaks in the spectrum of CET and other temperature records are close to harmonics of $T_{UN}$, Figure 2. The longest period component of CET resolvable by Fourier analysis occurs at 257 years making $3T_{UN}/2$ = 257.1 years an obvious choice as one of the centre periods for filtering. The spectral content in the broad band between 70 and 130 years in Figure 2 has no resolved peaks, however, the periods 68, 86, and 114 years cover the band reasonably well. The period 114 years was not used in equation 2 because the component at that period suffers a $\pi$ phase shift in the middle of the CET record and cannot be approximated as a single sinusoid. It is closely approximated by the term $0.075\cos(2\pi 1.5(t - 2019)/T_{UN}).\cos(2\pi(t - 2000)/3T_{UN})$ °C. However, when this term is added as an 11[th] term in equation 2 the correlation coefficient of projected CET with the CET record in Figure 8 improves only marginally, from 0.78 to 0.79. For this reason and the benefit of keeping equation 2 simple the term was not included.

**A 2. Phase shift between the 15-year component and cycle.** Of the cycles that are included in equation 2 the 15-year cycle suffers the largest phase shift relative to the corresponding component, about half a cycle over the 362-year record, Figure 17. This results in the short-term discrepancy between recorded and projected CET at around 2020 evident in Figures 6, 8, and 11.

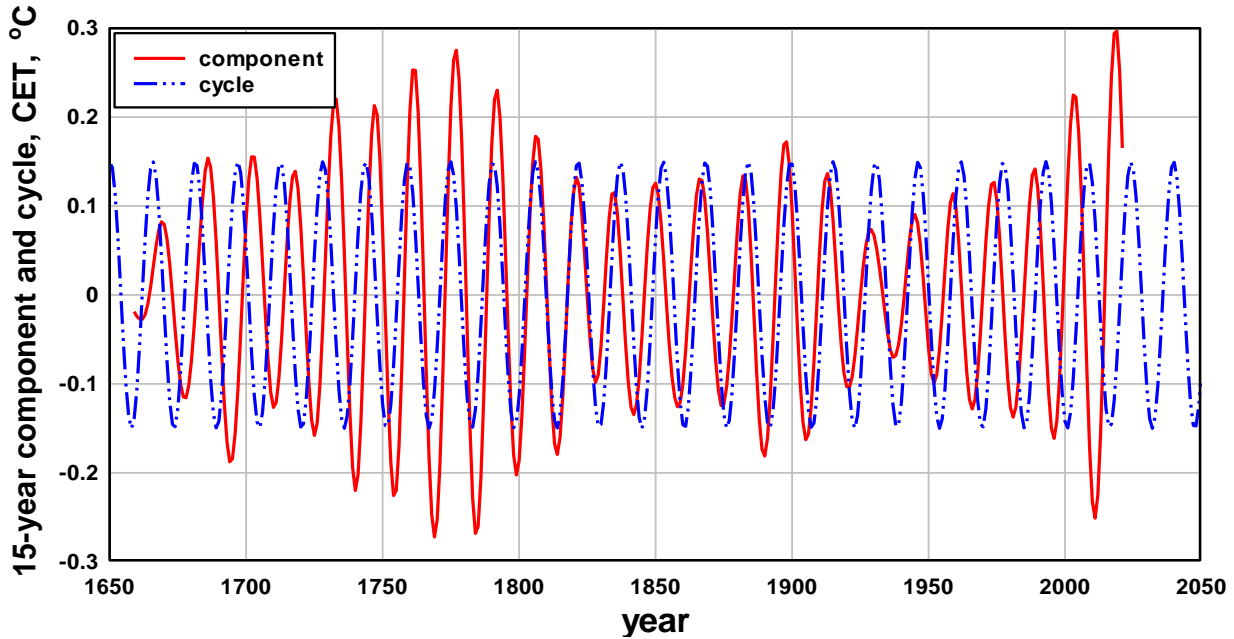

**Figure 17.** Compares the $T_{UN}/11$, 15.6-year, period component with the cycle approximation. The close to $\pi/2$ phase shift of the cycle relative to the component around year 2000 results in the short- term discrepancy between recorded CET and projected CET at this time.

**A 3. Basis for using $T_{UN}$ to decompose CET into cycles.** Decomposing the CET record into cycles with periods, T, given by $nT = mT_{UN}$ is based on evidence that the long-term periodicity of solar activity varies at harmonics of $T_{UN}$, McCracken et al (2014), and evidence that temperature on Earth varies with the variation in solar activity, (Usoskin et al 2005, Haigh 2007, Gray et al 2010, Bae et al 2022). The periods of the first ten harmonics of $T_{UN}$ are 85.7, 57.1, 42.8, 34.3, 28.5, 24.5, 21.4, 19.0, 17.1 and 15.6 years and four, 15, 24, 34, and 57 years, are evident in the CET spectrum, Figure 2. Other evidence that the harmonics of $T_{UN}$ are relevant to temperature variation is as follows: Previous Fourier analyses of the CET record are dominated by components with periods close to harmonics of $T_{UN}$. For example, the three long period cycles Baliunas et al (1997) identified in CET, were at periods of 102, 23.5, and 14.4 years, periods close to the first, sixth and tenth harmonics of $T_{UN}$. The five long period components in the Svalbard temperature record identified by Humlum et al (2011) are at periods 83, 62, 36, 26, and 16.8 years, close to the first, second, fourth, sixth, and ninth harmonics of $T_{UN}$ respectively. The five significant periods identified in six central European instrumental temperature records, Ludecke et al (2013) were at 248, 80, 61, 47 and 34 years; close to the 257, 86, 57, 43 and 34 year-periods used here. Mann et al (2021) identified oscillations in global mean temperature over the last millennium with periods at 66, 58, 43, 28, 24, 19, and 14 years all close to harmonics of $T_{UN}$ apart from 66 years. The period, 66 years, is close to period $2T_{UN}/5 = 68.5$ years that was used in this paper.

**Acknowledgment.** I acknowledge the use of CET data published by the Met Office, Hadley Centre for Climate Prediction and Research and aa index data from Geosciences Australia. I acknowledge useful comments by Don Field and Glen Johnston.